\documentstyle[twoside,graphicx]{article}

\catcode`\@=11
\long\def\@makefntext#1{
\protect\noindent \hbox to 3.2pt {\hskip-.9pt  
$^{{\eightrm\@thefnmark}}$\hfil}#1\hfill}               

\def\@makefnmark{\hbox to 0pt{$^{\@thefnmark}$\hss}}    
        
\def\ps@myheadings{\let\@mkboth\@gobbletwo
\def\@oddhead{\hbox{}
\rightmark\hfil\eightrm\thepage}   
\def\@oddfoot{}\def\@evenhead{\eightrm\thepage\hfil
\leftmark\hbox{}}\def\@evenfoot{}
\def\sectionmark##1{}\def\subsectionmark##1{}}

\oddsidemargin=\evensidemargin
\newcounter{sectionc}\newcounter{subsectionc}\newcounter{subsubsectionc}
\renewcommand{\section}[1] {\vspace{12pt}\addtocounter{sectionc}{1} 
\setcounter{subsectionc}{0}\setcounter{subsubsectionc}{0}\noindent 
        {\tenbf\thesectionc. #1}\par\vspace{5pt}}
\renewcommand{\subsection}[1] {\vspace{12pt}\addtocounter{subsectionc}{1} 
        \setcounter{subsubsectionc}{0}\noindent 
        {\bf\thesectionc.\thesubsectionc. {\kern1pt \bfit #1}}\par\vspace{5pt}}
\renewcommand{\subsubsection}[1] {\vspace{12pt}\addtocounter{subsubsectionc}{1}
        \noindent{\tenrm\thesectionc.\thesubsectionc.\thesubsubsectionc.
        {\kern1pt \tenit #1}}\par\vspace{5pt}}

\newcounter{appendixc}
\newcounter{subappendixc}[appendixc]
\newcounter{subsubappendixc}[subappendixc]
\renewcommand{\thesubappendixc}{\Alph{appendixc}.\arabic{subappendixc}}
\renewcommand{\thesubsubappendixc}
        {\Alph{appendixc}.\arabic{subappendixc}.\arabic{subsubappendixc}}

\renewcommand{\appendix}[1] {\vspace{12pt}
        \refstepcounter{appendixc}
        \setcounter{figure}{0}
        \setcounter{table}{0}
        \setcounter{lemma}{0}
        \setcounter{theorem}{0}
        \setcounter{corollary}{0}
        \setcounter{definition}{0}
        \setcounter{equation}{0}
        \renewcommand{\thefigure}{\Alph{appendixc}.\arabic{figure}}
        \renewcommand{\thetable}{\Alph{appendixc}.\arabic{table}}
        \renewcommand{\theappendixc}{\Alph{appendixc}}
        \renewcommand{\thelemma}{\Alph{appendixc}.\arabic{lemma}}
        \renewcommand{\thetheorem}{\Alph{appendixc}.\arabic{theorem}}
        \renewcommand{\thedefinition}{\Alph{appendixc}.\arabic{definition}}
        \renewcommand{\thecorollary}{\Alph{appendixc}.\arabic{corollary}}
        \renewcommand{\theequation}{\Alph{appendixc}.\arabic{equation}}
        \noindent{\tenbf Appendix \theappendixc #1}\par\vspace{5pt}}
\newcommand{\subappendix}[1] {\vspace{12pt}
        \refstepcounter{subappendixc}
        \noindent{\bf Appendix \thesubappendixc. {\kern1pt \bfit #1}}
        \par\vspace{5pt}}
\newcommand{\subsubappendix}[1] {\vspace{12pt}
        \refstepcounter{subsubappendixc}
        \noindent{\rm Appendix \thesubsubappendixc. {\kern1pt \tenit #1}}
        \par\vspace{5pt}}

\newcommand{\textlineskip}{\baselineskip=13pt}
\newcommand{\smalllineskip}{\baselineskip=10pt}

\def\eightcirc{
\begin{picture}(0,0)
\put(4.4,1.8){\circle{6.5}}
\end{picture}}
\def\eightcopyright{\eightcirc\kern2.7pt\hbox{\eightrm c}}

\def\abstracts#1#2#3{{
        \centering{\begin{minipage}{4.5in}\baselineskip=10pt\footnotesize
        \parindent=0pt #1\par 
        \parindent=15pt #2\par
        \parindent=15pt #3
        \end{minipage}}\par}} 

\renewenvironment{thebibliography}[1]
        {\frenchspacing
         \ninerm\baselineskip=11pt
         \begin{list}{\arabic{enumi}.}
        {\usecounter{enumi}\setlength{\parsep}{0pt}
         \setlength{\leftmargin 12.7pt}{\rightmargin 0pt} 
         \setlength{\itemsep}{0pt} \settowidth
        {\labelwidth}{#1.}\sloppy}}{\end{list}}

\newcounter{itemlistc}
\newcounter{romanlistc}
\newcounter{alphlistc}
\newcounter{arabiclistc}

\newcommand{\fcaption}[1]{
        \refstepcounter{figure}
        \setbox\@tempboxa = \hbox{\footnotesize Fig.~\thefigure. #1}
        \ifdim \wd\@tempboxa > 5in
           {\begin{center}
        \parbox{5in}{\footnotesize\smalllineskip Fig.~\thefigure. #1}
            \end{center}}
        \else
             {\begin{center}
             {\footnotesize Fig.~\thefigure. #1}
              \end{center}}
        \fi}

\newcommand{\tcaption}[1]{
        \refstepcounter{table}
        \setbox\@tempboxa = \hbox{\footnotesize Table~\thetable. #1}
        \ifdim \wd\@tempboxa > 5in
           {\begin{center}
        \parbox{5in}{\footnotesize\smalllineskip Table~\thetable. #1}
            \end{center}}
        \else
             {\begin{center}
             {\footnotesize Table~\thetable. #1}
              \end{center}}
        \fi}

\def\@citex[#1]#2{\if@filesw\immediate\write\@auxout
        {\string\citation{#2}}\fi
\def\@citea{}\@cite{\@for\@citeb:=#2\do
        {\@citea\def\@citea{,}\@ifundefined
        {b@\@citeb}{{\bf ?}\@warning
        {Citation `\@citeb' on page \thepage \space undefined}}
        {\csname b@\@citeb\endcsname}}}{#1}}

\newif\if@cghi
\def\cite{\@cghitrue\@ifnextchar [{\@tempswatrue
        \@citex}{\@tempswafalse\@citex[]}}
\def\citelow{\@cghifalse\@ifnextchar [{\@tempswatrue
        \@citex}{\@tempswafalse\@citex[]}}
\def\@cite#1#2{{$\null^{#1}$\if@tempswa\typeout
        {IJCGA warning: optional citation argument 
        ignored: `#2'} \fi}}

\def\pmb#1{\setbox0=\hbox{#1}
        \kern-.025em\copy0\kern-\wd0
        \kern.05em\copy0\kern-\wd0
        \kern-.025em\raise.0433em\box0}

\def\fnt#1#2{\footnotetext{\kern-.3em
        {$^{\mbox{\scriptsize #1}}$}{#2}}}

\font\tenrm=cmr10
\font\tenit=cmti10 
\font\tenbf=cmbx10
\font\bfit=cmbxti10 at 10pt
\font\ninerm=cmr9

\font\eightrm=cmr8

\newcommand{\rd}{{\mathrm{d}}}

\def\mathswitchr#1{\relax\ifmmode{\mathrm{#1}}\else$\mathrm{#1}$\fi}

\newcommand{\PW}{\mathswitchr W}

\newcommand{\Pe}{\mathswitchr e}

\newcommand{\Pep}{\mathswitchr {e^+}}
\newcommand{\Pem}{\mathswitchr {e^-}}

\def\mathswitch#1{\relax\ifmmode#1\else$#1$\fi}

\newcommand{\Me}{\mathswitch {m_\Pe}}

\def\solid{\raise.9mm\hbox{\protect\rule{1.1cm}{.2mm}}}
\def\dash{\raise.9mm\hbox{\protect\rule{2mm}{.2mm}}\hspace*{1mm}}

\def\ie{i.e.\ }

\newcommand{\sing}{{\mathrm{sing}}}
\newcommand{\finite}{{\mathrm{finite}}}

\newcommand{\DPA}{{\mathrm{DPA}}}

\newcommand{\Born}{{\mathrm{Born}}}

\newcommand{\virt}{{\mathrm{virt}}}
\newcommand{\real}{{\mathrm{real}}}

\newcommand{\eeWWffff}{\Pep\Pem\to\PW\PW\to 4f}
\newcommand{\eeffff}{\Pep\Pem\to 4f}
\newcommand{\eeffffg}{\eeffff\gamma}

\def\qed{\hbox{${\vcenter{\vbox{                        
   \hrule height 0.4pt\hbox{\vrule width 0.4pt height 6pt
   \kern5pt\vrule width 0.4pt}\hrule height 0.4pt}}}$}}

\begin{document}


\normalsize\textlineskip
\setcounter{page}{1}

\vspace*{-1.5cm}
\begin{flushright}
ER/40685/954 \\
UR-1616 \\
November 2000
\end{flushright}

\vspace*{0.1truein}

\centerline{\bf ELECTROWEAK RADIATIVE CORRECTIONS TO}
\vspace*{0.035truein}
\centerline{\bf OFF-SHELL W-PAIR PRODUCTION
\footnote{Talk given
at the DPF 2000 meeting, Columbus, OH, August 9-12, 2000.}
}
\vspace*{0.37truein}
\centerline{\footnotesize ANSGAR DENNER}
\vspace*{0.015truein}
\centerline{\footnotesize\it Paul Scherrer Institut}
\baselineskip=10pt
\centerline{\footnotesize\it CH-5232 Villigen PSI, Switzerland}
\vspace*{10pt}
\centerline{\footnotesize STEFAN DITTMAIER}
\vspace*{0.015truein}
\centerline{\footnotesize\it Theoretische Physik, Universit\"at Bielefeld}
\baselineskip=10pt
\centerline{\footnotesize\it D-33615 Bielefeld, Germany}
\vspace*{10pt}
\centerline{\footnotesize MARKUS ROTH}
\vspace*{0.015truein}
\centerline{\footnotesize\it Institut f\"ur Theoretische Physik,
Universit\"at Leipzig}
\baselineskip=10pt
\centerline{\footnotesize\it D-04109 Leipzig, Germany}
\vspace*{10pt}
\centerline{\footnotesize DOREEN WACKEROTH} 
\vspace*{0.015truein}
\centerline{\footnotesize\it Department of Physics and Astronomy, 
University of Rochester}
\baselineskip=10pt
\centerline{\footnotesize\it Rochester, New York 14627, USA}

\vspace*{0.21truein}
\abstracts{We briefly describe the RacoonWW approach to calculate 
radiative corrections to $\eeWWffff$ and 
present numerical results for the total W-pair production
cross section at LEP2.}{}{}

\vspace*{1pt}\textlineskip      
\section{Introduction}          
\vspace*{-0.5pt}
\noindent
At LEP2, for the first time, a precise direct measurement of the
triple gauge-boson couplings $(\gamma,\mathrm{Z})$WW is
performed\cite{lep2exp}, allowing to test the non-abelian structure of
the Standard Model of electroweak interactions (SM). Moreover, a
measurement of the W-boson mass at LEP2 with a precision of 35 MeV is
within reach\cite{lep2exp}, which will not only provide a further
precisely known SM input parameter but will also help to considerably
improve the present indirect bounds on the Higgs-boson
mass\cite{Abbaneo:2000nr}.  To match the experimental precision at
LEP2 and a future Linear $\Pep \Pem$ Collider, the theoretical
predictions for the cross sections to $\eeWWffff$ need to be of ${\cal
  O}(1\%)$ accuracy and better\cite{lep2mcws}.  This requires to go
beyond the inclusion of universal corrections such as the running of
the electromagnetic coupling, corrections to the $\rho$ parameter, the
Coulomb singularity, and leading photonic corrections, as described in
Ref.\ref{been}. The theoretical uncertainty of predictions which only
include universal radiative corrections is estimated\cite{lep2rep} to
be of ${\cal O}(1$--$2\%)$ for the total W-pair production cross
section, $\sigma_{\mathrm{WW}}$, at LEP2 energies and is expected to
be even larger for differential cross sections.  So far, 
no complete calculation of the electroweak ${\cal O}(\alpha)$
corrections to the processes $\eeffff$ exists.  Such a calculation is not
only enormously complex but also poses severe theoretical problems
with gauge invariance connected to the instability of the gauge
bosons.  Nevertheless, there are efforts in this direction\cite{Vi98}.
At present, all predictions for $\eeWWffff$, which go beyond universal
radiative corrections, rely on a Double-Pole-Approximation (DPA):
radiative corrections are taken into account only to those
contributions that are enhanced by two resonant W propagators.
Different versions of the DPA are implemented in the Monte Carlo
generators RacoonWW\cite{Denner:2000bj} and YFSWW3\cite{yfsww}, and
the results of a tuned comparison at LEP2 and LC energies are
presented in Refs.\ref{grune} and \ref{mrst}, respectively.  As a
result of this comparison, supported by a tuned comparison with the
semi-analytic calculation of Ref.\ref{bbc} and a detailed study of the
intrinsic uncertainty of the DPA with {\sc RacoonWW}, the four-fermion
working group of the LEP2MC workshop\cite{lep2mcws} assigned a
theoretical uncertainty to the predictions for $\sigma_{\mathrm{WW}}$
by {\sc RacoonWW} and YFSWW3 of $0.4 \%$ at energies between 200 and
500 GeV. 
In the following we give a brief description of the {\sc RacoonWW} approach.

\section{\boldmath{Electroweak Radiative Corrections to $\eeWWffff$}}
\noindent
The complete ${\cal O}(\alpha)$ corrections to off-shell W-pair
production in DPA as implemented in {\sc RacoonWW} can
be written as follows\cite{Denner:2000bj}:
\begin{equation}\label{eq:crosssection}
\rd \sigma_{\mathrm{WW}} =
\rd \sigma_{\Born}^{\eeffff}+
\rd \sigma_{\virt,\finite,\DPA}^{\eeWWffff}
+\rd\sigma_{\virt+\real,\sing}^{\eeffff}
+\rd\sigma_{\finite}^{\eeffffg}.
\end{equation}
Here $\rd\sigma_{\Born}^{\eeffff}$ is the full lowest-order cross
section to $\eeffff$ and $\rd\sigma^{\eeffffg}$, which describes real
photon radiation away from IR and collinear singularities, is the full
lowest-order cross section to $\eeffffg$ as described in
Ref.\ref{paperg}.  The cross section for real photon radiation is
rendered finite in two different ways: by using phase-space
slicing or alternatively a subtraction method\cite{subtr}.  The exact
$4f$ and $4f+\gamma$ phase spaces for massless fermions are used
throughout. The IR-finite sum of virtual, soft and collinear photonic
one-loop corrections is denoted by $\rd
\sigma_{\virt+\real,\sing}^{\eeffff}$. Apart from finite terms, it
contains only collinear singularities associated with the initial
state, \ie leading logarithms of the form $\ln(s/\Me^2)$, at least for
inclusive enough observables.  Since the contribution
$\rd\sigma_{\virt+\real,\sing}^{\eeffff}$ is not treated in DPA, those
logarithms are treated in our approach exactly.  In {\sc RacoonWW},
the DPA is only applied to the finite part of the virtual corrections,
denoted by $\rd\sigma_{\virt,\finite,\DPA}^{\eeWWffff}$ including the
full set of factorizable and non-factorizable virtual ${\cal
  O}(\alpha)$ corrections. The factorizable contribution comprises
those virtual one-loop corrections that can either be assigned to the
W-pair production or to the W-decay subprocesses and, thus can be
deduced from the known results for on-shell W-pair production and W
decay.  The non-factorizable corrections connect the production and
decay processes. In DPA, they comprise all doubly-resonant
contributions that are not already included in the factorizable
corrections.  Beyond ${\cal O}(\alpha)$, {\sc RacoonWW} includes
soft-photon exponentiation and leading higher-order initial-state
radiation up to ${\cal O}(\alpha^3)$ via the structure-function
approach\cite{lep2rep}. The leading higher-order effects from $\Delta
\rho$ and $\Delta \alpha$ are taken into account by using the
$G_{\mu}$ scheme.  QCD corrections are taken into account
either by a multiplicative factor $(1+\alpha_{\mathrm s}/ \pi)$ for
each hadronically decaying W boson or as complete QCD corrections
\looseness -1
to the CC03 diagrams.

\begin{figure}[htbp]
\vspace*{13pt}
\includegraphics[height=6.8cm]{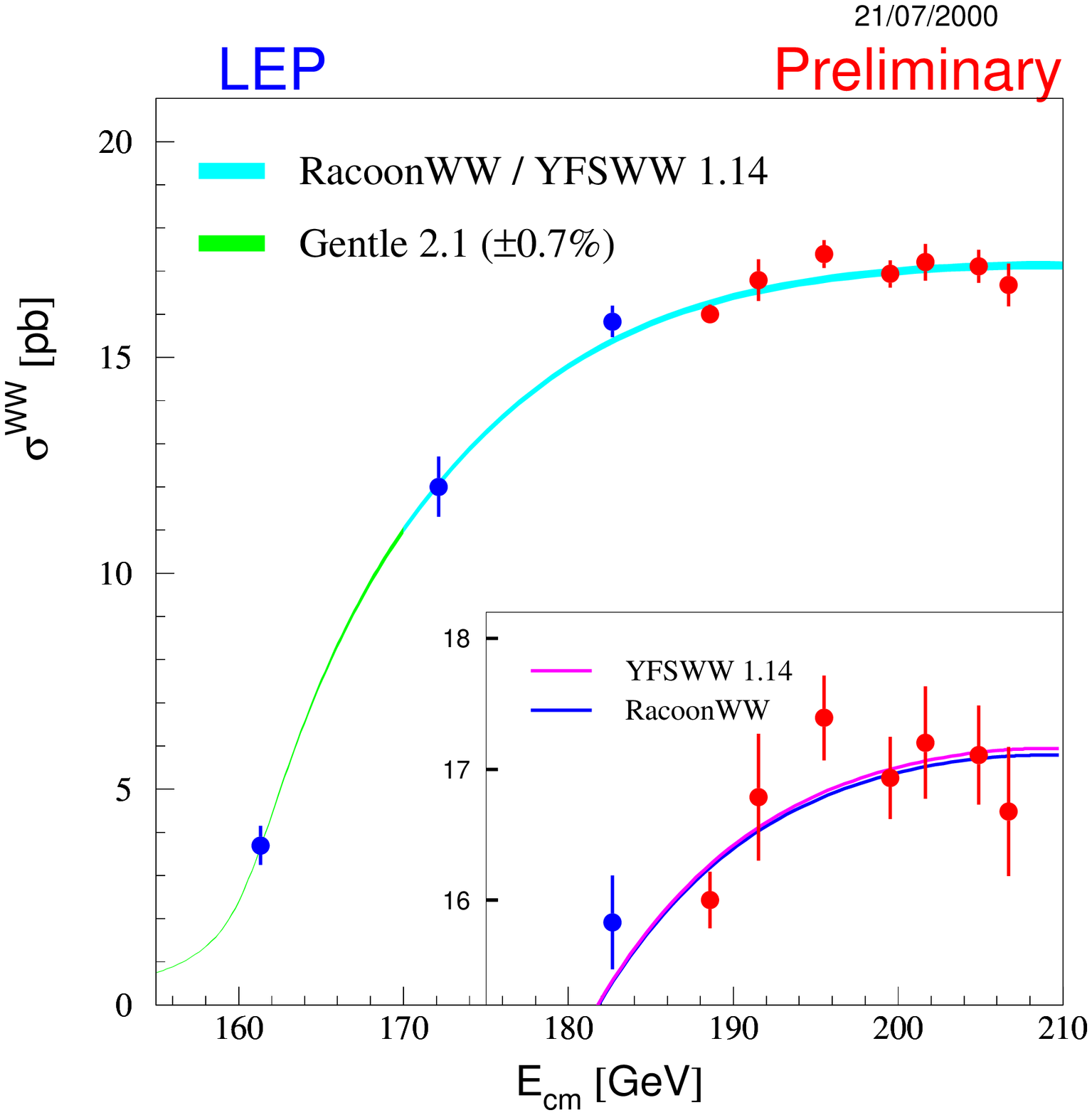}
\includegraphics[height=6.8cm,width=5.5cm]{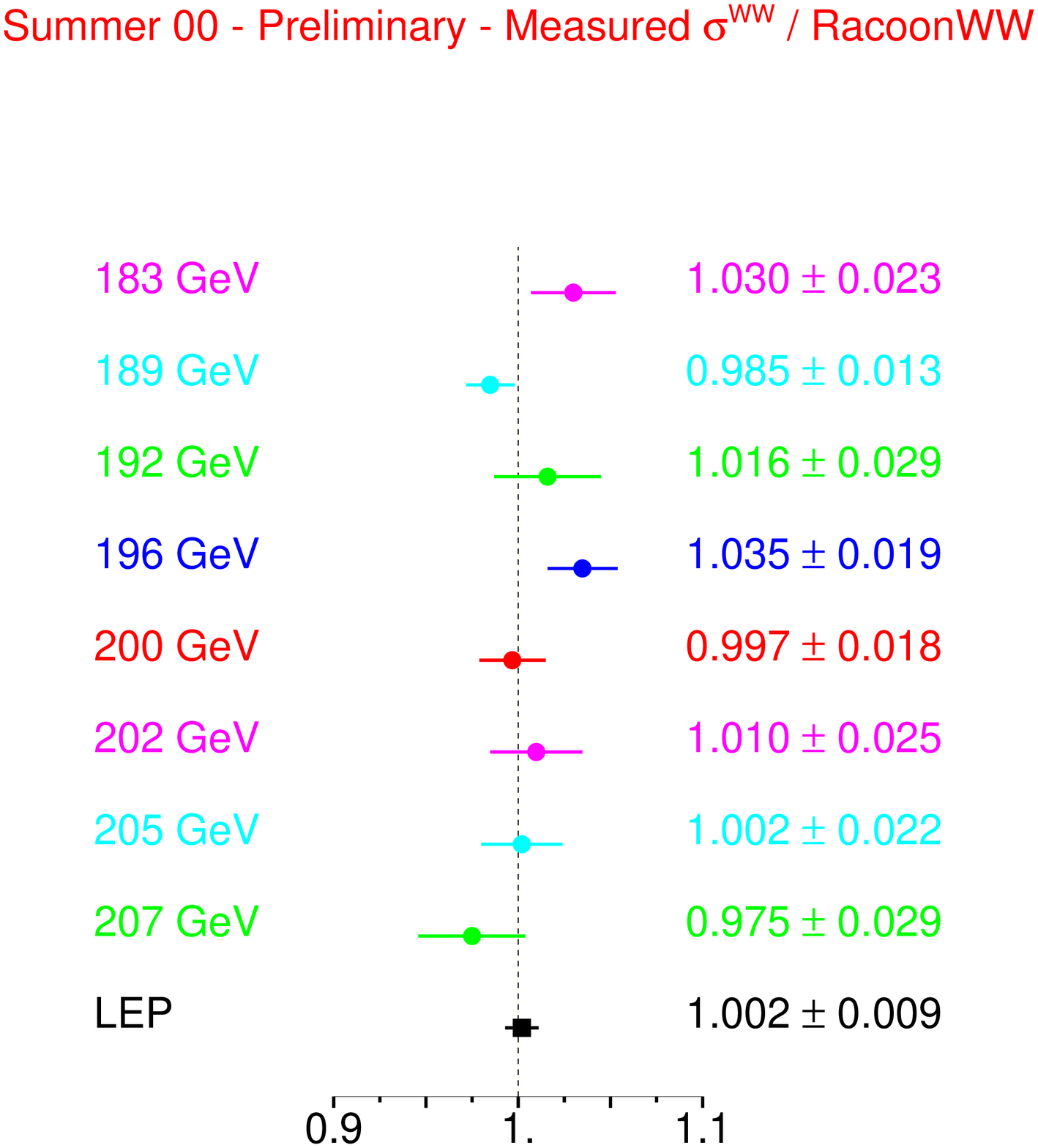}
\vspace*{13pt}
\fcaption{The total W-pair production cross section and the ratio
of the data over the RacoonWW prediction at LEP2, as given by the 
LEPEWWG\cite{Abbaneo:2000nr}.}\label{fig:wackone}
\end{figure}

Numerical results from {\sc RacoonWW} at LEP2 energies for some
interesting observables, such as invariant-mass distributions of the W
decay products and angular distributions, are provided in
Refs.\ref{grune},\ref{paper},\ref{letter}. 
In Fig.\ref{fig:wackone}, we show the
comparison of the {\sc RacoonWW} prediction for the W-pair-production
cross section with LEP2 data. From a comparison of {\sc RacoonWW} with
GENTLEv2.00\cite{gentle}, for instance, the neglect of non-universal
${\cal O}(\alpha)$ corrections is known to induce a $2.5\%$
shift\cite{lep2mcws} to higher values of $\sigma_{\mathrm{WW}}$ which
is disfavoured by the data.  The good agreement of LEP2 data
with the state-of-the-art Monte Carlo generators {\sc RacoonWW} and
YFSWW3 reveals evidence of non-leading electroweak 
corrections beyond the level of universal effects.
\looseness -1

\end{document}